\documentstyle[aps,graphics]{revtex}
\begin{document}
\draft
\title{Weak-Coupling-Like Time Evolution of Driven Four-Level Systems in the
Strong-Coupling Regime}
\author{Vicente Delgado and J. M. Gomez Llorente}
\address{Departamento de F\'{\i }sica Fundamental II,\\
Universidad de La Laguna, 38205-La Laguna, Tenerife, Spain}
\maketitle

\begin{abstract}
It is shown analytically that there exists a natural basis in terms of which
the nonperturbative time evolution of an important class of driven
four-level systems in the strong-coupling regime decouples and essentially
reduces to the corresponding time evolution in the weak-field regime,
exhibiting simple Rabi oscillations between the different relevant quantum
states. The predictions of the model are corroborated by an exact numerical
calculation.
\end{abstract}


\section{Introduction}

The dynamical behavior of quantum systems driven by external time-dependent
fields has attracted considerable interest in recent years due, in part, to
the great variety of phenomena that have been theoretically predicted and
experimentally observed when the system conditions are conveniently chosen 
\cite{Grif,Ficek,Joachain}. For instance, in the field of quantum optics the
quantum interference effects induced by the coherent external fields can
lead to phenomena such as coherent population trapping \cite{Ari1} (even in
the nonperturbative regime \cite{VDBJM}), electromagnetically induced
transparency \cite{Harr1}, or lasing without inversion \cite{Koch1}. In
atomic systems an external laser field can induce interesting processes such
as harmonic generation \cite{Kraus1} and multiphoton excitation and
ionization \cite{Gravila}.

The theoretical treatment of a quantum system exposed to a strong
time-dependent field requires specific nonperturbative methods. A first
comprehensive theoretical study of the effects of a strong oscillating field
on a two-level quantum system was carried out by Autler and Townes \cite
{Autler}{, who making use of Floquet's theorem \cite{Floquet} derived a
solution in terms of infinite continued fractions} to investigate the effect
of an rf field on the $J=2\rightarrow 1\;l$-type doublet microwave
absorption lines of molecules of gaseous OCS, obtaining good agreement with
the experimental results. In another important paper Shirley \cite{Shir1}
used also the {Floquet's theorem} to develop a general formalism for
treating periodically driven quantum systems. Using this formalism, which
replaces the solution of the time-dependent Schr\"{o}dinger equation with
the solution of a time-independent Schr\"{o}dinger equation represented by
an infinite matrix, he obtained closed expressions for time-average
resonance transition probabilities of a strongly-driven two-level system.
More recently, a variety of approaches have been proposed to deal
analytically with strongly driven two-level systems \cite
{Plata,Bavli,Zhao,Frasca,Delgado,Barata,Sac}. Three- and four-level systems
driven by intense laser fields has also been treated analytically \cite
{VDBJM,Plata2}.

In the numerical description of realistic multi-level atoms and molecules in
intense laser fields the Floquet theory and, more recently, the
R-matrix-Floquet approach \cite{Burke1}, have also proved to be particularly
useful. These formalisms have been used in studies of atomic spectroscopy 
\cite{Zhang}, laser-assisted electron-atom scattering \cite{Potv}, harmonic
generation \cite{Geba}, periodically kicked Rydberg atoms \cite{Yosh1} and
multiphoton excitation and ionization of atoms and molecules \cite
{Sal,Chu,Leas,Shak,Dorr}.

In this work we are interested in an important class of driven four-level
systems. Specifically, we consider a four-level system consisting of two
doublets (see Fig. 1). This system has been previously studied in the
context of coherent population transfer \cite{Shore} and tunneling dynamics 
\cite{Plata2}. In the present work we will show that there exists a natural
basis in terms of which the nonperturbative time evolution of the system in
the strong-coupling regime decouples and essentially reduces to the
corresponding time evolution in the weak-field regime.

The splittings of the two lower states ($|1\rangle $, $|2\rangle $) and the
two upper states ($|3\rangle $, $|4\rangle $) will be denoted as $\Delta
_{0}^{\prime }$ and $\Delta _{0}^{\prime \prime }$, respectively. These
splittings are much smaller than the separation $\Delta $ between the
doublets. Such a level configuration is commonly encountered in quantum
double-well potentials, which in turn are relevant in the description of
numerous processes in molecular and solid-state systems. For instance, this
model can describe the tunneling dynamics of the inversion mode of the
ammonia molecule \cite{Grif,Hund}, intermolecular proton transfer processes 
\cite{Oppen}, or the effect of a driving laser field on the tunneling
dynamics of low-lying electrons in quantum semiconductor heterostructures 
\cite{Holt1,Ferr1}.

The external periodic field (of amplitude ${\bf E}$ and frequency $\omega $)
will induce transitions between states $|1\rangle \leftrightarrow |2\rangle $%
, $|3\rangle \leftrightarrow |4\rangle $, $|1\rangle \leftrightarrow
|4\rangle $, and $|2\rangle \leftrightarrow |3\rangle $, with corresponding
coupling constants $\Omega _{12}$, $\Omega _{34}$, $\Omega _{14}$, and $%
\Omega _{23}$, where $\Omega _{ij}\equiv {\bf E\,\mu }_{ij}$ and ${\bf \mu }%
_{ij}$ is the dipole matrix element between states $|i\rangle
\leftrightarrow |j\rangle $. The Hamiltonian of the system reads

%%%%%%%%%%%%%% Fig. 1 %%%%%%%%%%%%%%%

\begin{figure}{\par\centering \resizebox{4.cm}{!}{\rotatebox{0}
{\includegraphics{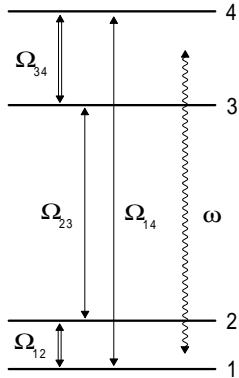}}} \par}
\caption{\small
Energy diagram and coupling constants of the four-level system considered
in the present work.
}\end{figure}

\begin{eqnarray}
H &=&\sum_{i=1}^{4}E_{i}\sigma _{ii}-\Omega _{12}\cos \left( \omega t\right)
\left( \sigma _{12}+\sigma _{21}\right) -\Omega _{34}\cos \left( \omega
t\right) \left( \sigma _{34}+\sigma _{43}\right)  \nonumber \\
&&-\Omega _{14}\cos \left( \omega t\right) \left( \sigma _{14}+\sigma
_{41}\right) -\Omega _{23}\cos \left( \omega t\right) \left( \sigma
_{23}+\sigma _{32}\right) ,  \label{ec1.1}
\end{eqnarray}
where $\sigma _{ij}\equiv |i\rangle \langle j|$ is the transition operator, $%
E_{i}$ is the energy of state $|i\rangle $ in the absence of the periodic
force, and we take $\hbar \equiv 1$ throughout the paper.

We shall assume that, as is usually the case, the dipole matrix elements $%
{\bf \mu }_{ij}$ between states lying within a given doublet are much larger
than the corresponding dipole matrix elements connecting states lying in
different doublets. Under these circumstances, the states within a doublet
are much more strongly coupled by the external field than those lying in
different doublets, i.e., $\Omega _{12}$, $\Omega _{34}\gg \Omega _{14}$, $%
\Omega _{23}$. If the driving field is quasiresonant with the allowed
transitions between the lower and upper doublets ($\omega \approx \Delta $)
and weak enough so that {\em any} coupling constant $\Omega _{ij}$ is much
smaller than the frequency $\omega $ (weak-coupling regime), the
contribution of the far-off-resonant transitions $|1\rangle \leftrightarrow
|2\rangle $ and $|3\rangle \leftrightarrow |4\rangle $ turns out to be
negligible. Under these conditions one can invoke the rotating wave
approximation (RWA) and the dynamical evolution of the system becomes
governed by a Hamiltonian which, in the rotating frame, takes the simple
form $H^{\prime }=H_{14}^{\prime }+H_{23}^{\prime }$ with

\begin{equation}
H_{14}^{\prime }\equiv E_{1}\sigma _{11}+\left( E_{4}-\omega \right) \sigma
_{44}-\frac{\Omega _{14}}{2}\left( \sigma _{14}+\sigma _{41}\right) ,
\label{ec1.2}
\end{equation}
and a similar expression for $H_{23}^{\prime }$ replacing $1\rightarrow 2$
and $4\rightarrow 3$. Thus, in the weak-field regime, the time evolution
consists of usual Rabi oscillations between $|1\rangle \leftrightarrow
|4\rangle $ and $|2\rangle \leftrightarrow |3\rangle $. As the strength of
the external field increases, the contribution of the strong
nonresonant-transitions $|1\rangle \leftrightarrow |2\rangle $ and $%
|3\rangle \leftrightarrow |4\rangle $ becomes increasingly important so
that, eventually, the system enters an interesting regime in which the
intra-doublet transitions become strong while the corresponding
inter-doublet transitions remain weak. This is a nonperturbative
strong-coupling regime where the RWA is no longer valid and one has to deal
with two weak and two strong transitions. Under these circumstances all the
states become coupled and the dynamical evolution becomes, in general,
rather involved. As we will show, there exists, however, a natural basis in
terms of which the time evolution of the system is essentially the same in
both the perturbative and nonperturbative regimes. This basis thus provides
a unified description of the weak- and strong-coupling regimes.

Our approach is not directly based on Floquet theory, rather it relies on a
suitable time-dependent unitary transformation which allows the
(intra-doublet) strong contributions to be conveniently absorbed into
renormalized physical parameters. However, a connection can be established
between these two approaches. Floquet states and quasienergies, which are,
respectively, the eigenstates and eigenvalues of the hermitian operator $%
(H-i\partial /\partial t)$, become in the time-independent case
indistinguishable from the usual stationary states and energies. Thus, if
one performs, as we shall do, a unitary transformation $U(t)$ to a rotating
frame in which the transformed Hamiltonian becomes, to a good approximation,
time-independent, then finding the Floquet states reduces to the
straightforward task of diagonalizing the rotated Hamiltonian and
transforming back to the original frame by applying $U^{+}(t)$. As we shall
see later on, the natural basis which provides a unified description of the
weak- and strong-coupling regimes is nothing but the basis of Floquet states
associated with a zeroth-order Hamiltonian obtained from the original
Hamiltonian (\ref{ec1.1}) by decoupling the two doublets, i.e., by taking $%
\Omega _{14}$, $\Omega _{23}\rightarrow 0$.

\section{Analytic model}

We start by performing the time-dependent unitary transformation

\begin{equation}
U(t)=\exp i\int_{0}^{t}\left[ \sum_{i=1}^{4}\frac{1}{2}\left(
E_{1}+E_{2}\right) \sigma _{ii}+\omega \left( \sigma _{33}+\sigma
_{44}\right) -\Omega _{12}\cos \left( \omega t^{\prime }\right) \left(
\sigma _{12}+\sigma _{21}\right) -\Omega _{34}\cos \left( \omega t^{\prime
}\right) \left( \sigma _{34}+\sigma _{43}\right) \right] dt^{\prime },
\label{ec1.3}
\end{equation}
This transformation, which in particular translates the zero of energy to
the point $\left( E_{1}+E_{2}\right) /2$, enables us to absorb the most
rapidly oscillating terms of the Hamiltonian (\ref{ec1.1}) and leads, after
some lengthy algebra, to the following rotated Hamiltonian:

\begin{equation}
\begin{array}{c}
H^{\prime }=\left( \Delta _{0}^{\prime }/2\right) \left\{ \cos \left[ 2\phi
^{\prime }(t)\right] \left( \sigma _{22}-\sigma _{11}\right) \right. +\left.
i\sin \left[ 2\phi ^{\prime }(t)\right] \left( \sigma _{21}-\sigma
_{12}\right) \right\} \\ 
+\left( \Delta _{0}^{\prime \prime }/2\right) \left\{ \cos \left[ 2\phi
^{\prime \prime }(t)\right] \left( \sigma _{44}-\sigma _{33}\right) \right.
+\left. i\sin \left[ 2\phi ^{\prime \prime }(t)\right] \left( \sigma
_{43}-\sigma _{34}\right) \right\} +(\Delta -\omega )\left( \sigma
_{33}+\sigma _{44}\right) \\ 
-\frac{1}{2}\left( \Omega _{14}+\Omega _{23}\right) \cos \left( \omega
t\right) \left\{ \,e^{-i\omega t}\left( \cos \left[ \phi _{-}(t)\right]
\left( \sigma _{23}+\sigma _{14}\right) -i\sin \left[ \phi _{-}(t)\right]
\left( \sigma _{13}+\sigma _{24}\right) \right) +{\rm h.c.}\right\} \\ 
-\frac{1}{2}\left( \Omega _{23}-\Omega _{14}\right) \cos \left( \omega
t\right) \left\{ \,e^{-i\omega t}\left( \cos \left[ \phi _{+}(t)\right]
\left( \sigma _{23}-\sigma _{14}\right) -i\sin \left[ \phi _{+}(t)\right]
\left( \sigma _{13}-\sigma _{24}\right) \right) +{\rm h.c.}\right\} ,
\end{array}
\label{ec1.4}
\end{equation}
with 
\begin{mathletters}
\begin{equation}
\phi ^{\prime }(t)=\left( \Omega _{12}/\omega \right) \sin \left( \omega
t\right),  \label{ec1.5a}
\end{equation}
\begin{equation}
\phi ^{\prime \prime }(t)=\left( \Omega _{34}/\omega \right) \sin \left(
\omega t\right),  \label{ec1.5b}
\end{equation}
\begin{equation}
\phi _{\pm }(t)=\phi ^{\prime}(t)\pm \phi ^{\prime \prime }(t).
\label{ec1.5c}
\end{equation}
Next, we express the time-dependent coefficients of $H^{\prime }$ as a
series of Bessel functions $J_{n}$ by using the expansions \cite{Abramo} 
\end{mathletters}
\begin{mathletters}
\begin{equation}
\cos \left[ \zeta \sin \left( \omega t\right) \right] =J_{0}\left( \zeta
\right) +2\sum_{n=1}^{+\infty }J_{2n}\left( \zeta \right) \cos \left(
2n\omega t\right),  \label{ec1.6a}
\end{equation}
\begin{equation}
\sin \left[ \zeta \sin \left( \omega t\right) \right] =2\sum_{n=0}^{+\infty
}J_{2n+1}\left( \zeta \right) \sin \left[ (2n+1)\omega t\right].
\label{ec1.6b}
\end{equation}
This enables us to write the Hamiltonian (\ref{ec1.4}) as a sum of a
dominant constant contribution $H_{0}^{\prime }$ and an oscillating
time-dependent part $H_{1}^{\prime }(t)$, with

\end{mathletters}
\begin{equation}
\begin{array}{c}
H_{0}^{\prime }=\frac{1}{2}\Delta _{0}^{\prime {\rm R}}\left( \sigma
_{22}-\sigma _{11}\right) +\frac{1}{2}\Delta _{0}^{\prime \prime {\rm R}%
}\left( \sigma _{44}-\sigma _{33}\right) +(\Delta -\omega )\left( \sigma
_{33}+\sigma _{44}\right) \\ 
-\frac{1}{2}\Omega _{14}^{{\rm R}}\left( \sigma _{14}+\sigma _{41}\right) -%
\frac{1}{2}\Omega _{23}^{{\rm R}}\left( \sigma _{23}+\sigma _{32}\right) ,
\end{array}
\label{ec1.7}
\end{equation}

\begin{equation}
\begin{array}{c}
H_{1}^{\prime }(t)=\Delta _{0}^{\prime }\sum_{n=1}^{+\infty }J_{2n}\left(
\zeta ^{\prime }\right) \cos \left( 2n\omega t\right) \left( \sigma
_{22}-\sigma _{11}\right) +i\Delta _{0}^{\prime }\sum_{n=0}^{+\infty
}J_{2n+1}\left( \zeta ^{\prime }\right) \sin \left[ (2n+1)\omega t\right]
\left( \sigma _{21}-\sigma _{12}\right) \\ 
+\Delta _{0}^{\prime \prime }\sum_{n=1}^{+\infty }J_{2n}\left( \zeta
^{\prime \prime }\right) \cos \left( 2n\omega t\right) \left( \sigma
_{44}-\sigma _{33}\right) +i\Delta _{0}^{\prime \prime }\sum_{n=0}^{+\infty
}J_{2n+1}\left( \zeta ^{\prime \prime }\right) \sin \left[ (2n+1)\omega
t\right] \left( \sigma _{43}-\sigma _{34}\right) \\ 
-\frac{1}{2}\left( \Omega _{23}+\Omega _{14}\right) \left[ \chi _{{\rm c}%
}(\zeta _{-})\left( \sigma _{23}+\sigma _{14}\right) -i\chi _{{\rm s}}(\zeta
_{-})\left( \sigma _{13}+\sigma _{24}\right) +{\rm h.c.}\right] \\ 
-\frac{1}{2}\left( \Omega _{23}-\Omega _{14}\right) \left[ \chi _{{\rm c}%
}(\zeta _{+})\left( \sigma _{23}-\sigma _{14}\right) -i\chi _{{\rm s}}(\zeta
_{+})\left( \sigma _{13}-\sigma _{24}\right) +{\rm h.c.}\right] ,
\end{array}
\label{ec1.8}
\end{equation}
where $\zeta ^{\prime }\equiv 2\Omega _{12}/\omega $, $\zeta ^{\prime \prime
}\equiv 2\Omega _{34}/\omega $ and $\zeta _{\pm }\equiv \frac{1}{2} \left(
\zeta^{\prime }\pm \zeta ^{\prime \prime }\right) $. The renormalized
splittings $\Delta _{0}^{\prime {\rm R}}$, $\Delta _{0}^{\prime \prime {\rm R%
}}$ and Rabi frequencies $\Omega _{14}^{{\rm R}}$, $\Omega _{23}^{{\rm R}}$
are field-dependent quantities defined as $\Delta _{0}^{\prime {\rm R}%
}=\Delta _{0}^{\prime }J_{0}\left( 2\Omega _{12}/\omega \right) $, $\Delta
_{0}^{\prime \prime {\rm R}}=\Delta _{0}^{\prime \prime }J_{0}\left( 2\Omega
_{34}/\omega \right) $, $\Omega _{14}^{{\rm R}}=\Omega _{+}^{{\rm R}}$ and $%
\Omega _{23}^{{\rm R}}=\Omega _{-}^{{\rm R}}$ with

\begin{equation}
\Omega _{\pm }^{{\rm R}}=\omega \left\{ \frac{\Omega _{14}+\Omega _{23}}{%
\Omega _{12}-\Omega _{34}}J_{1}\left( \frac{\Omega _{12}-\Omega _{34}}{%
\omega }\right) \pm \frac{\Omega _{14}-\Omega _{23}}{\Omega _{12}+\Omega
_{34}}J_{1}\left( \frac{\Omega _{12}+\Omega _{34}}{\omega } \right) \right\}
,  \label{ec1.9}
\end{equation}
and the coefficients $\chi _{{\rm c}}(\zeta )$ and $\chi _{{\rm s}}(\zeta )$
are defined by

\begin{mathletters}
\begin{equation}  \label{ec1.10}
\chi _{{\rm c}}(\zeta )=\frac 12e^{-2i\omega t}\left\{ J_0\left( \zeta
\right) +e^{-2i\omega t}J_2\left( \zeta \right) \right\}
+\sum_{n=1}^{+\infty }\left\{ 1+\left( 1-\delta _{n1}\right) e^{-2i\omega
t}\right\} J_{2n}\left( \zeta \right) \cos \left( 2n\omega t\right) ,
\end{equation}

\begin{equation}
\chi _{{\rm s}}(\zeta )=\left( 1+e^{-2i\omega t}\right) \sum_{n=0}^{+\infty
}J_{2n+1}\left( \zeta \right) \sin \left[ (2n+1)\omega t\right] ,
\label{ec1.11}
\end{equation}
where $\delta _{n1}$ is the Kronecker delta.

The important point is that the contribution of the oscillating Hamiltonian $%
H_{1}^{\prime }(t)$ to the dynamical evolution of the system becomes
negligible and can be safely neglected under rather general conditions. To
see this, we write the evolution operator associated with $H^{\prime
}(t)=H_{0}^{\prime }+H_{1}^{\prime }(t)$ as the perturbative expansion

\end{mathletters}
\begin{equation}
U^{\prime }(t,0)=e^{-iH_{0}^{\prime }t}\left\{ {\bf 1}-i\int_{0}^{t}dt^{%
\prime }e^{iH_{0}^{\prime }t^{\prime }}H_{1}^{\prime }(t^{\prime
})e^{-iH_{0}^{\prime }t^{\prime }}+\ldots \right\} .  \label{ec1.13}
\end{equation}
It can be easily seen that the integral on the right hand side of (\ref
{ec1.13}) is a sum of terms of order $\Delta _{0}^{\prime }/\omega $, $%
\Delta _{0}^{\prime \prime }/\omega $, $\Omega _{14}/\omega $, and $\Omega
_{23}/\omega $. Thus, for a driving field quasiresonant with the transitions
between the lower and upper doublets ($\omega \approx \Delta \gg \Delta
_{0}^{\prime },\Delta _{0}^{\prime \prime }$) and weak enough so that the
Rabi frequencies of the weak transitions ($\Omega _{14},\Omega _{23}$)
remain small compared to $\omega $, the contribution of $H_{1}^{\prime }(t)$
can be legitimately neglected. This approximation is applicable regardless
of the value of the coupling constants $\Omega _{12}$ and $\Omega _{34}$
and, therefore, is valid in both the perturbative and nonperturbative
regimes. In particular, in the weak-field regime it leads to the same
results as the usual RWA and, consequently, can be considered as a
nonperturbative generalization of the latter. Under the above conditions,
the dynamical evolution becomes governed by the Hamiltonian $H_{0}^{\prime }$
which, by defining renormalized energies $E_{1}^{{\rm R}}=-\Delta
_{0}^{\prime {\rm R}}/2$, $E_{2}^{{\rm R}}=\Delta _{0}^{\prime {\rm R}}/2$, $%
E_{3}^{{\rm R}}=\Delta -\Delta _{0}^{\prime \prime {\rm R}}/2$, and $E_{4}^{%
{\rm R}}=\Delta +\Delta _{0}^{\prime \prime {\rm R}}/2$, it takes the same
form as the weak-field Hamiltonian previously considered. Specifically, one
obtains $H^{\prime }=H_{14}^{\prime }+H_{23}^{\prime }$ with

\begin{equation}
H_{14}^{\prime }\equiv E_{1}^{{\rm R}}\sigma _{11}+\left( E_{4}^{{\rm R}%
}-\omega \right) \sigma _{44}-\frac{\Omega _{14}^{{\rm R}}}{2}\left( \sigma
_{14}+\sigma _{41}\right) ,  \label{ec1.14}
\end{equation}
and a similar expression for $H_{23}^{\prime }$ replacing $1\rightarrow 2$
and $4\rightarrow 3$. The Schr\"{o}dinger equation associated with $%
H^{\prime }$ can be now readily solved analytically to obtain the
nonperturbative general solution in the rotating frame

\begin{equation}
|\psi ^{\prime}(t)\rangle =\sum_{i=1}^{4}c_{i}^{\prime }(t)|i\rangle ,
\label{ec1.15a}
\end{equation}
with probability amplitudes $c_{i}^{\prime }(t)$ given by

\begin{mathletters}
\begin{equation}  \label{ec1.16}
c_1^{\prime }(t)=\left\{ c_1^{\prime }(0)\cos \left( \frac{\overline{\Omega }%
_{14}^{{\rm R}}}2t\right) +\frac i{\overline{\Omega }_{14}^{{\rm R}}}\left(
c_1^{\prime }(0)\,\delta _{14}^{{\rm R}}\right. \right. +\left. \left.
c_4^{\prime }(0)\Omega _{14}^{{\rm R}}\right) \sin \left( \frac{\overline{%
\Omega }_{14}^{{\rm R}}}2t\right) \right\} e^{-i\left( \delta _{14}^{{\rm R}%
}/2+E_1^{{\rm R}}\right) t} ,
\end{equation}

\begin{equation}  \label{ec1.17}
c_2^{\prime }(t)=\left\{ c_2^{\prime }(0)\cos \left( \frac{\overline{\Omega }%
_{23}^{{\rm R}}}2t\right) +\frac i{\overline{\Omega }_{23}^{{\rm R}}}\left(
c_2^{\prime }(0)\,\delta _{23}^{{\rm R}}\right. \right. +\left. \left.
c_3^{\prime }(0)\Omega _{23}^{{\rm R}}\right) \sin \left( \frac{\overline{%
\Omega }_{23}^{{\rm R}}}2t\right) \right\} e^{-i\left( \delta _{23}^{{\rm R}%
}/2+E_2^{{\rm R}}\right) t} ,
\end{equation}

\begin{equation}  \label{ec1.18}
c_3^{\prime }(t)=\left\{ c_3^{\prime }(0)\cos \left( \frac{\overline{\Omega }%
_{23}^{{\rm R}}}2t\right) -\frac i{\overline{\Omega }_{23}^{{\rm R}}}\left(
c_3^{\prime }(0)\,\delta _{23}^{{\rm R}}\right. \right. -\left. \left.
c_2^{\prime }(0)\Omega _{23}^{{\rm R}}\right) \sin \left( \frac{\overline{%
\Omega }_{23}^{{\rm R}}}2t\right) \right\} e^{i\left( \delta _{23}^{{\rm R}%
}/2-E_3^{{\rm R}}+\omega \right) t} ,
\end{equation}

\begin{equation}
c_{4}^{\prime }(t)=\left\{ c_{4}^{\prime }(0)\cos \left( \frac{\overline{%
\Omega }_{14}^{{\rm R}}}{2}t\right) -\frac{i}{\overline{\Omega }_{14}^{{\rm R%
}}}\left( c_{4}^{\prime }(0)\,\delta _{14}^{{\rm R}}\right. \right. -\left.
\left. c_{1}^{\prime }(0)\Omega _{14}^{{\rm R}}\right) \sin \left( \frac{%
\overline{\Omega }_{14}^{{\rm R}}}{2}t\right) \right\} e^{i\left( \delta
_{14}^{{\rm R}}/2-E_{4}^{{\rm R}}+\omega \right) t} ,  \label{ec1.19}
\end{equation}
where we have defined field-dependent renormalized detunings $\delta _{14}^{%
{\rm R}}=E_{4}^{{\rm R}}-E_{1}^{{\rm R}}-\omega $ and $\delta _{23}^{{\rm R}%
}=E_{3}^{{\rm R}}-E_{2}^{{\rm R}}-\omega $, and renormalized
generalized-Rabi frequencies $\overline{\Omega }_{14}^{{\rm R}}=\sqrt{\left(
\Omega _{14}^{{\rm R}}\right) ^{2}+\left( \delta _{14}^{{\rm R}}\right) ^{2}}
$ and $\overline{\Omega }_{23}^{{\rm R}}=\sqrt{\left( \Omega _{23}^{{\rm R}%
}\right) ^{2}+\left( \delta _{23}^{{\rm R}}\right) ^{2}}$. The interesting
point is that while the system dynamics in the strong-field regime is in
general rather complicated, when viewed from the rotating frame it becomes
essentially the same as that of the weak-field regime. The same result holds
true in the original nonrotating frame by proper choice of the relevant
basis. Indeed, by transforming back one obtains

\end{mathletters}
\begin{equation}  \label{ec1.19b}
|\psi (t)\rangle=U^{+}(t)|\psi ^{\prime }(t)\rangle =
\sum_{i=1}^{4}c_{i}^{\prime}(t)U^{+}(t)|i\rangle ,
\end{equation}
and from this general solution one immediately sees that the probability
amplitudes associated with the field-dependent states $|i^{\prime
}(t)\rangle \equiv U^{+}(t)|i\rangle $ are precisely those given in Eqs. (%
\ref{ec1.16}--\ref{ec1.19}). It is therefore clear that the renormalized $%
|i^{\prime }(t)\rangle $ states constitute the natural basis to analyze the
time evolution of the system. In fact, when the system dynamics is analyzed
in terms of such states the nonperturbative effects induced by the strong
driving field can be absorbed into a redefinition of the relevant energies
and Rabi frequencies in such a way that the system evolves obeying the same
Hamiltonian in both the perturbative and nonperturbative regimes.

In terms of the original basis, the states $\left\{ |i^{\prime }(t)\rangle
\right\} $ take the form

\begin{mathletters}
\begin{equation}
|1^{\prime }(t)\rangle =\cos \phi ^{\prime }(t)|1\rangle +i\sin \phi
^{\prime }(t)|2\rangle ,  \label{ec1.20}
\end{equation}

\begin{equation}
|2^{\prime }(t)\rangle =i\sin \phi ^{\prime }(t)|1\rangle +\cos \phi
^{\prime }(t)|2\rangle ,  \label{ec1.21}
\end{equation}

\begin{equation}
|3^{\prime }(t)\rangle =e^{-i\omega t}\left[ \cos \phi ^{\prime \prime
}(t)|3\rangle +i\sin \phi ^{\prime \prime }(t)|4\rangle \right] ,
\label{ec1.22}
\end{equation}

\begin{equation}
|4^{\prime }(t)\rangle =e^{-i\omega t}\left[ i\sin \phi ^{\prime \prime
}(t)|3\rangle +\cos \phi ^{\prime \prime }(t)|4\rangle \right] .
\label{ec1.23}
\end{equation}
These states constitute a basis of the extended Hilbert space of $t$%
-periodic state vectors \cite{Sambe}. In fact, as already mentioned, they
are the Floquet states associated with the zeroth-order Hamiltonian obtained
from the original Hamiltonian (\ref{ec1.1}) by decoupling the two doublets,
i.e., by taking $\Omega _{14}$, $\Omega _{23}\rightarrow 0$. This follows
from the fact that, in such a case, the states $\left\{ |i\rangle \right\} $
become the eigenstates of the rotated Hamiltonian $H^{\prime
}=H_{14}^{\prime }+H_{23}^{\prime }$ (see Eq. (\ref{ec1.14}) and below).

Note that in the weak-field regime one has $\phi ^{\prime }(t)$, $\phi
^{\prime \prime }(t)\ll 1$ for any $t$ and as a consequence $|i^{\prime
}(t)\rangle \rightarrow |i\rangle $, so that the renormalized basis becomes
indistinguishable from the original one. Similarly, taking into account that 
$J_{0}\left( x\right) \rightarrow 1$ and $J_{1}\left( x\right) /x\rightarrow
1/2$ as $x\rightarrow 0$ it follows that in such regime the renormalized
energies and Rabi frequencies approach their corresponding bare values, so
that, in the weak-field regime, the above formulation simply reduces to the
usual one. In the strong-field regime, however, the time evolution of the
different bare states becomes strongly coupled by the driving field and as a
consequence it can be rather involved and very different from that occurring
in the weak-field regime. In contrast, the time evolution of the
renormalized states remains always as simple as in the weak-field regime,
consisting of Rabi oscillations between $|1^{\prime }(t)\rangle
\leftrightarrow |4^{\prime }(t)\rangle $ and $|2^{\prime }(t)\rangle
\leftrightarrow |3^{\prime }(t)\rangle $.

\section{Numerical results}

To verify the predictions of the above analytic model, next we perform an
exact numerical calculation. We consider a quantum particle in a quartic
double-well potential driven by an external periodic field of frequency $%
\omega $ (see Fig. 2). Since this potential approaches an infinite value at
large distances, $x\rightarrow \pm \infty ,$ it only admits bound
eigenstates \cite{Schiff}. Consequently, there is no continuum spectrum and
such a model is only adequate for describing physical systems at energies
well below the continuum threshold.

%%%%%%%%%%%%%% Fig. 2 %%%%%%%%%%%%%%%

\begin{figure}{\par\centering \resizebox{5.cm}{!}{\rotatebox{0}
{\includegraphics{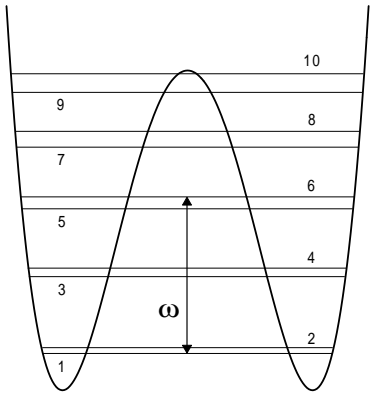}}} \par}
\caption{\small
Energy diagram of the lowest lying eigenstates of a quartic double-well 
potential with $D=4$. The splittings of the doublets have been exaggerated 
for clarity. %
}\end{figure}

Using convenient dimensionless variables, the corresponding Hamiltonian can
be cast in the form \cite{Gros1}

\end{mathletters}
\begin{equation}
H=\frac{\hat{p}^{2}}{2}-\frac{\hat{x}^{2}}{4}+\frac{\hat{x}^{4}}{64D}%
-\lambda \hat{x}\cos \omega t.  \label{ec1.24}
\end{equation}

The dimensionless parameter $D$ determines the barrier height and
corresponds, approximately, to the number of doublets below the top of the
barrier. In the present study we take $D=4$. The frequency of the external
field has been tuned to the transitions between the first and third
doublets. Specifically, we have taken $\omega =(E_{6}-E_{1})$ with $E_{i}$
being the energy levels in the absence of driving field. The dimensionless
field intensity $\lambda $, on the other hand, has been chosen to satisfy
the strong-coupling condition $\Omega _{12}/\omega =1$, where $\Omega
_{12}=\lambda \langle 1\left| \hat{x}\right| 2\rangle $ (see below).

To establish a more clear connection with the formalism of the previous
Sections, it is convenient to rewrite the Hamiltonian (\ref{ec1.24}) in
terms of a basis set $\left\{ |i\rangle \right\} $ of eigenstates of the
quartic oscillator. We have obtained these eigenstates numerically, by
diagonalization in another truncated basis set $\left\{ |\varphi _{n}\rangle
\right\} $ of harmonic oscillator wave functions with a conveniently
optimized frequency, following the procedure of Ref.\cite{Ballen}. In this
way, one gains a complete knowledge of the states $|i\rangle
=\sum_{n}\langle \varphi _{n}|i\rangle |\varphi _{n}\rangle $ by determining
the numerical coefficients $\langle \varphi _{n}|i\rangle .$

In terms of the quartic-oscillator eigenstates, the Hamiltonian (\ref{ec1.24}%
) takes the form

\begin{equation}
H=\sum_{i,j}|i\rangle \langle i|H|j\rangle \langle j|=\sum_{i}E_{i}\sigma
_{ii}-\lambda \cos \omega t\sum_{i,j}\langle i|\hat{x}|j\rangle \sigma _{ij}.
\label{ec1.25}
\end{equation}
where $\sigma _{ij}\equiv |i\rangle \langle j|$ and $E_{i}$ is the energy of
state $|i\rangle $ in the absence of the periodic force. Introducing the
notation $\Omega _{ij}\equiv \lambda \langle i\left| \hat{x}\right| j\rangle
=\lambda \langle j\left| \hat{x}\right| i\rangle $ for the coupling
constants, the connection with the formalism developed previously should now
be evident.

Note that, since $\hat{x}$ is an odd operator and the parity of the
quartic-oscillator eigenstate $|i\rangle $ is $(-1)^{i-1}$ (with $%
i=1,2,3,\ldots $), transitions between $|i\rangle \leftrightarrow |j\rangle $
are allowed only if $(i-j)$ is odd.

For a sufficiently weak driving field $\lambda $, any coupling constant $%
\Omega _{ij}$ will be much smaller than $\omega $ and the system will evolve
in a weak-coupling regime where all of the allowed transitions are weak.
Conversely, for a sufficiently intense driving field we would have $\Omega
_{ij}\gg \omega $ for any $i,j$ and the system would evolve in a
nonperturbative strong-coupling regime where all of the allowed transitions
are strong. However, as mentioned in the Introduction, in between these two
limiting cases there exists an interesting nonperturbative strong-coupling
regime in which the intra-doublet transitions become strong while the
corresponding inter-doublet transitions remain weak. This is so due to the
fact that, in this kind of systems, the dipole matrix elements $\mu
_{ij}\sim \langle i\left| \hat{x}\right| j\rangle $ between states within a
given doublet turn out to be much larger than the dipole matrix elements
connecting states lying in different doublets. On the other hand, under the
above conditions, since inter-doublet transitions are weak, contributions
coming from off-resonant doublets will be negligible, so that one expects
the quartic oscillator to behave, to a good approximation, as an effective
four-level system. Note, finally, that the nonperturbative regime in which 
we are interested lies within the range of applicability of the analytic
formalism of Sec. II [see below Eq. (\ref{ec1.13})].

We have solved numerically the time-dependent Schr\"{o}dinger equation
corresponding to the Hamiltonian (\ref{ec1.25}) by expanding its solution in
the basis set of eigenstates of the quartic oscillator, and have considered
as many states in the truncated basis sets so as to guarantee well-converged
results. Specifically, for the physical parameters considered above, the 20
lowest-lying levels of the quartic oscillator have been included, which are
more than enough to guarantee convergence. As we shall see, under the above
conditions, the dynamical evolution of the system can be described, to a
good approximation, by a four-level model.

In what follows we shall denote the two states of the upper doublet as $%
|3\rangle $ and $|4\rangle $, in accordance with the notation used in the
four-level analytical model developed in the previous Section. Figure 3
shows the time evolution of the populations $\left| \langle i|\psi
(t)\rangle \right| ^{2}$ of the bare states $|i\rangle $ ($i=1,2,3,4$) for a
system prepared in $t=0$ in the ground state. The curves plotted correspond
to the numerical results obtained by solving the Schr\"{o}dinger equation
with the Hamiltonian (\ref{ec1.24}). In Fig. 4 we show the corresponding
theoretical prediction, obtained from the analytic general solution $|\psi
(t)\rangle $ given by Eq. (\ref{ec1.19b}). It is important to note that the
numerical results, unlike the analytical ones, include the contribution from
all of the energy levels (and not only the contribution from the most
relevant four levels). In fact, the slight discrepancy between Figs. 3 and 4
is due entirely to this circumstance, as demonstrates the fact that both
analytical and numerical results become indistinguishable when the numerical
problem is also restricted to the four most relevant levels.

%%%%%%%%%%%%%% Fig. 3 %%%%%%%%%%%%%%%

\begin{figure}{\par\centering \resizebox{14.cm}{!}{\rotatebox{0}
{\includegraphics{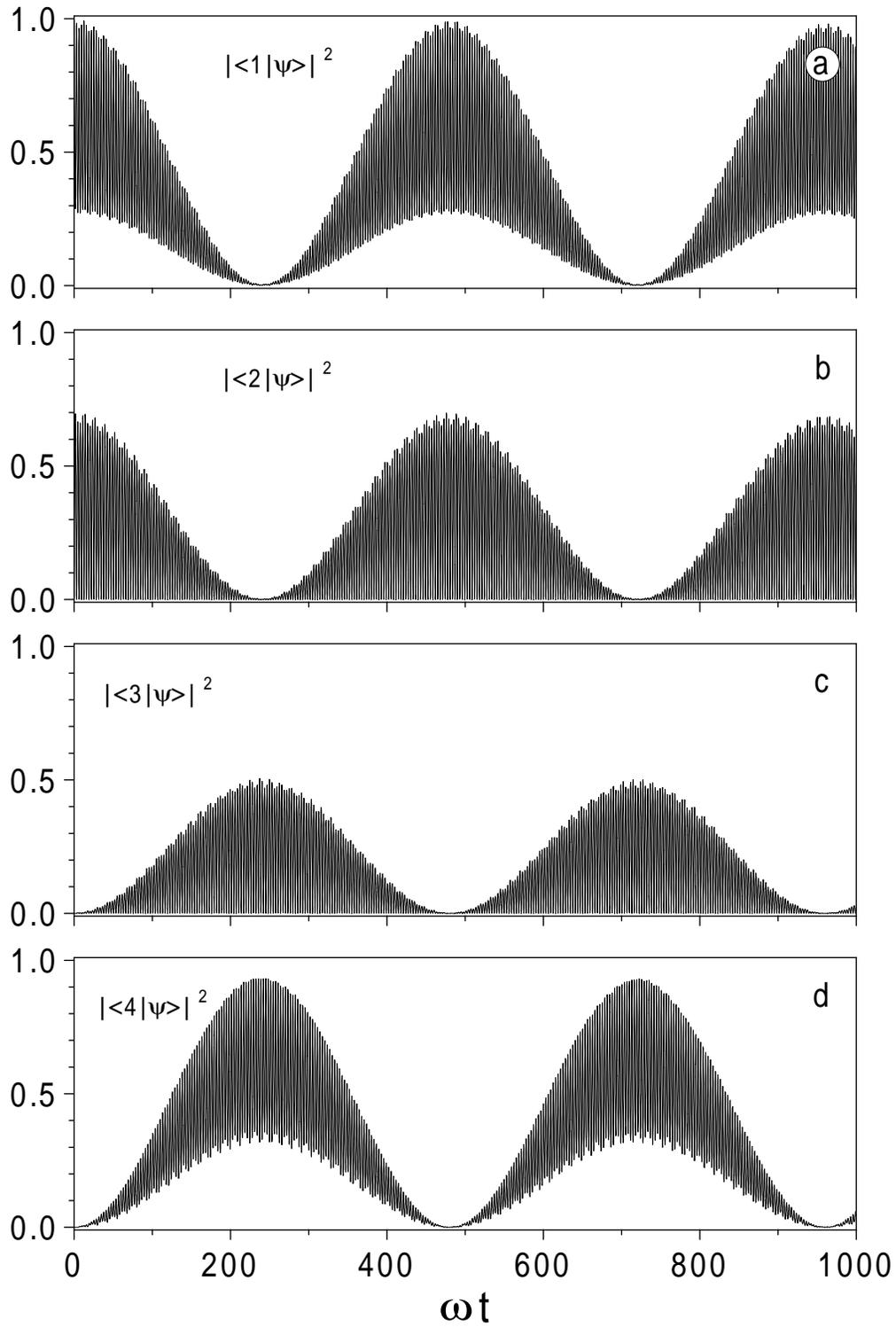}}} \par}
\caption{\small 
Dimensionless time evolution of the populations of the bare states 
$|i\rangle $ for a system initially prepared in the ground state. %
}
\end{figure}

%%%%%%%%%%%%%% Fig. 4 %%%%%%%%%%%%%%%

\begin{figure}{\par\centering \resizebox{14.cm}{!}{\rotatebox{0}
{\includegraphics{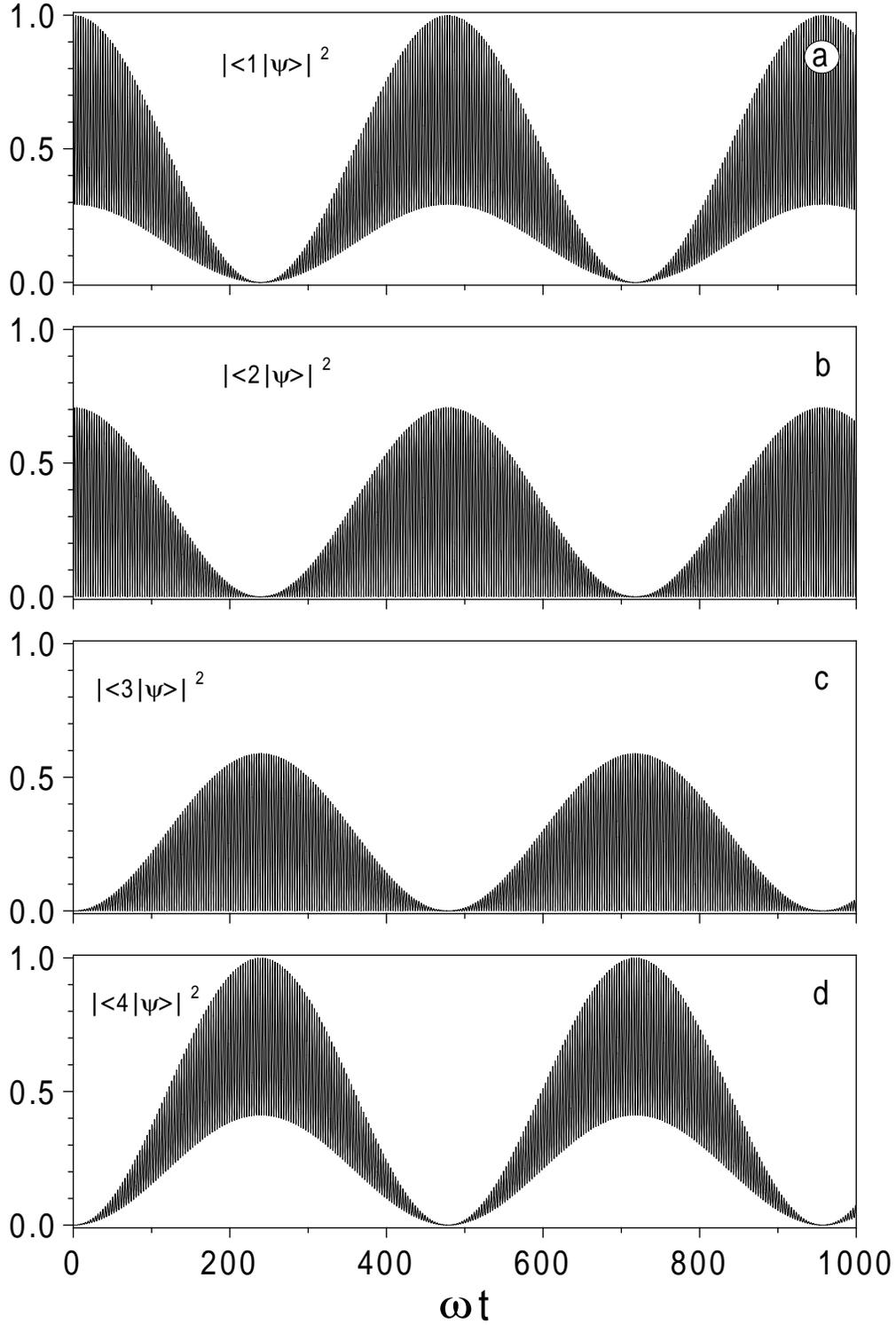}}} \par}
\caption{\small
Theoretical prediction corresponding to Fig. 3. %
}\end{figure}

Figures 3 and 4 show that under the action of the strong external field all
of the bare states become highly populated and their time evolution couples
in such a way that the population dynamics turns out to be quite different
from the simple Rabi oscillations occurring in the weak-field regime. In
contrast, as Fig. 5 reflects, the populations of the renormalized states $%
|i^{\prime }(t)\rangle $ evolve in time exhibiting the usual Rabi
oscillations of the weak-field regime. Solid lines in this figure correspond
to exact numerical results whereas dashed lines correspond to the analytical
results obtained from Eqs. (\ref{ec1.16}--\ref{ec1.19}). As before, the
small difference between analytical and numerical results originates from
corrections to the four-level approximation. Indeed, for the high field
intensity considered above, the contribution of the second and forth
doublets to the dynamical evolution of the system, although small, it is not
completely negligible. In fact, by monitoring the different numerical
populations it can be seen that a small proportion of the populations of the
first and third doublets is rapidly transferred to their corresponding
adjacent doublets, giving rise to the rapid oscillations that appear
superimposed to the usual Rabi oscillations in Fig. 5. When the numerical
problem is restricted to the four most relevant levels this population
transfer vanishes and, as already mentioned, both analytical and numerical
results become indistinguishable. Since the contribution of level $|i\rangle 
$ to the dynamical evolution of level $|j\rangle $ is proportional to $%
\Omega _{ij}/\delta _{ij}$ (where $\Omega _{ij}$ is the field-dependent
coupling constant between $|i\rangle $ and $|j\rangle $, and $\delta _{ij}$
is the detuning of the corresponding transition \cite{Cohen}), one expects a
better agreement between analytical and numerical results for smaller field
intensities. This is indeed the case as can be appreciated from Fig. 6. This
figure shows the population dynamics of the renormalized states for an
external field of the same frequency as before but a smaller field
intensity, which now satisfies the strong-coupling conditions $\Omega
_{12}/\omega =0.75$ (Fig. 6a) and $\Omega _{12}/\omega =0.5$ (Fig. 6b). As
expected, as the intensity of the external field decreases the four-level
approximation becomes more and more exact.

%%%%%%%%%%%%%% Fig. 5 %%%%%%%%%%%%%%%

\begin{figure}{\par\centering \resizebox{17.cm}{!}{\rotatebox{0}
{\includegraphics{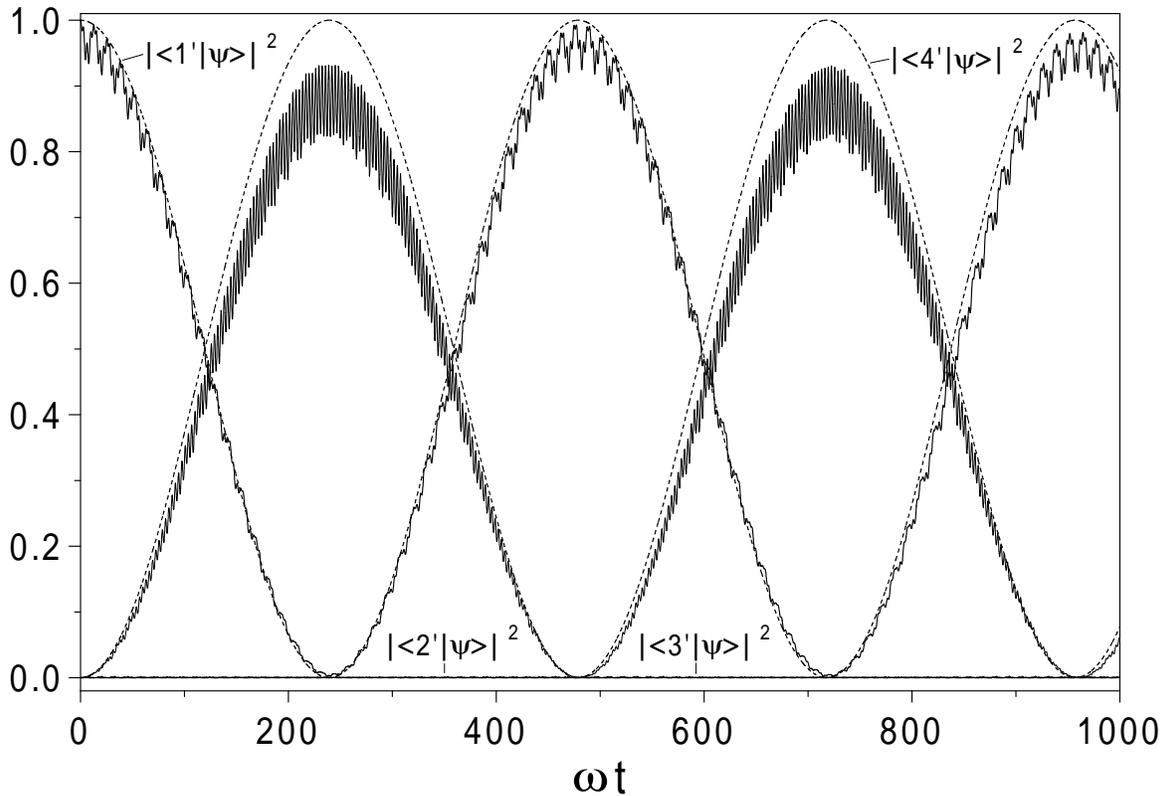}}} \par}
\caption{\small
Dimensionless time evolution of the populations of the renormalized 
states $|i^{\prime }(t)\rangle $ for a system initially prepared in 
the ground state.  
Solid lines are exact numerical results and dashed lines are the 
analytical results. %
}\end{figure}

Figures 5 and 6 show that, for the above initial conditions, states $%
|2^{\prime }(t)\rangle $ and $|3^{\prime }(t)\rangle $ remain unpopulated
while the system population undergoes Rabi oscillations between the
renormalized states $|1^{\prime }(t)\rangle $ and $|4^{\prime }(t)\rangle $,
and this occurs in both the weak- and strong-field regimes.

%%%%%%%%%%%%%% Fig. 6 %%%%%%%%%%%%%%%

\begin{figure}{\par\centering \resizebox{14.cm}{!}{\rotatebox{0}
{\includegraphics{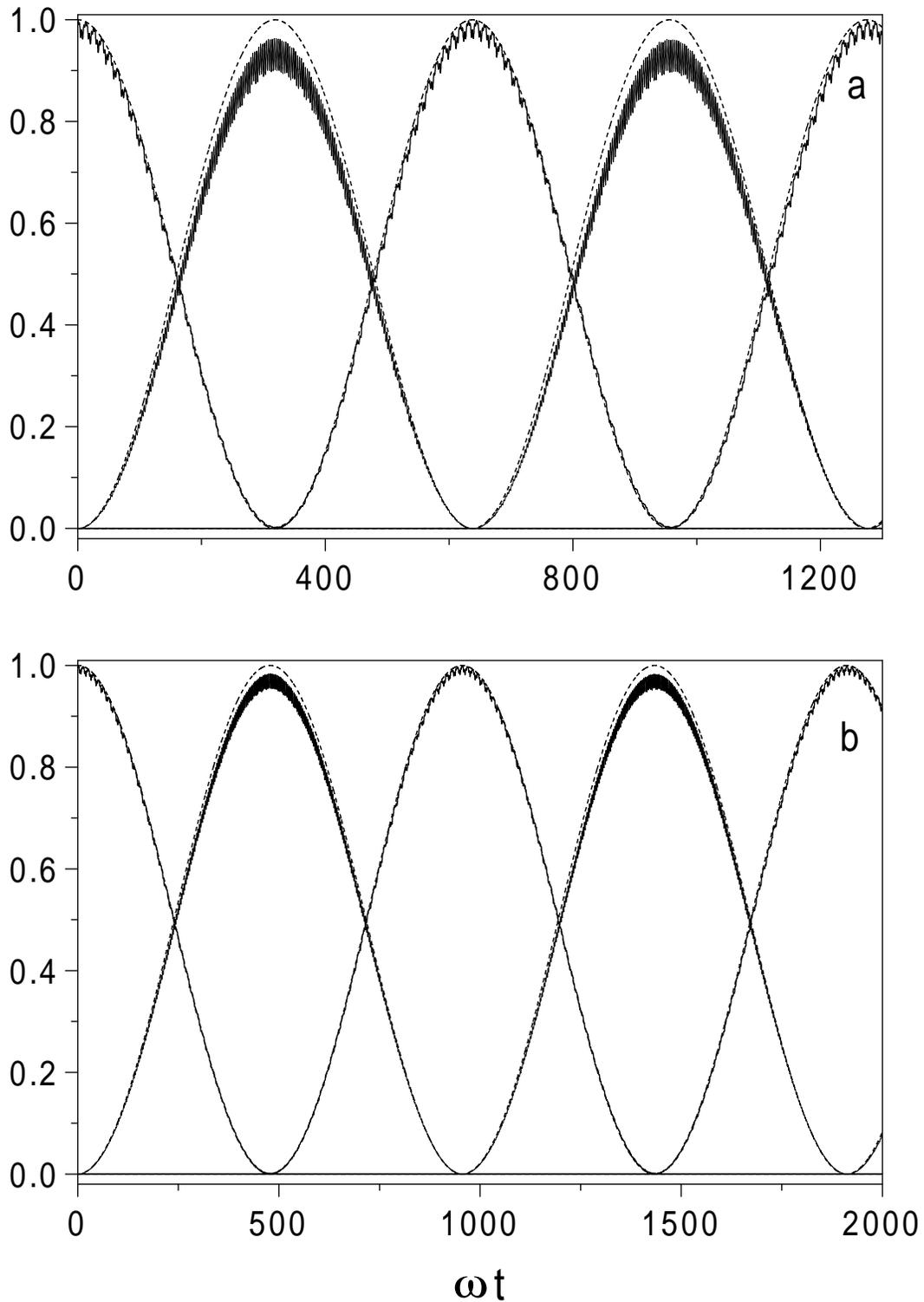}}} \par}
\caption{\small
The same curves as in Fig. 5 for an external field of the same frequency 
as before but having now an intensity satisfying the conditions (a) 
$\Omega_{12}/\omega =0.75$ and (b) $\Omega _{12}/\omega =0.5$. %
}\end{figure}

\section{Conclusion}

In the nonperturbative regime the dynamical behavior of driven quantum
systems becomes, in general, rather involved. In this paper we have
considered an important class of driven four-level systems which are
relevant in the description of numerous processes in molecular and
solid-state systems, and we have shown that their nonperturbative time
evolution, when analyzed in terms of a natural basis of renormalized states,
essentially reduces to the corresponding time evolution in the weak-field
regime, exhibiting simple Rabi oscillations between the different relevant
quantum states.

Such renormalized basis enables one to absorb the nonperturbative effects
induced by the strong driving field into a redefinition of the relevant
energies and Rabi frequencies in such a way that the system evolves obeying
the same Hamiltonian in the perturbative and nonperturbative regimes. This
basis thus provides a unified description valid in both the weak- and
strong-coupling regimes. In particular, in the weak-field regime, the
renormalized basis becomes indistinguishable from the original one and the
renormalized energies and Rabi frequencies approach their corresponding bare
values, so that, in this regime, our formulation leads to the same results
as the usual RWA and thus can be considered as a nonperturbative
generalization of the latter.

This work has been supported by Ministerio de Ciencia y Tecnolog\'{i}a and
FEDER under Grant No. BFM2001-3343.

%\end{mathletters}

\end{document}